\documentclass[journal]{IEEEtran}
\ifCLASSINFOpdf
\usepackage[pdftex]{graphicx}
\graphicspath{{../pdf/}{../jpeg/}}
\DeclareGraphicsExtensions{.pdf,.jpeg,.png}
\else
\usepackage[dvips]{graphicx}
\graphicspath{{../eps/}}
\DeclareGraphicsExtensions{.eps}
\fi
\usepackage{graphicx}
\usepackage{graphics}
\usepackage{epsfig}
\usepackage{epstopdf}

\usepackage{epstopdf}
\usepackage{stfloats}
\usepackage[cmex10]{amsmath}
\usepackage{algorithmic}
\usepackage{array}
\usepackage{mdwmath}
\usepackage{mdwtab}
\usepackage{graphicx}
\usepackage{subfigure}
\usepackage{color}
\usepackage{amsfonts,amssymb}
\usepackage{hyperref} 
\usepackage{lineno} 
\usepackage{multirow}
\usepackage{diagbox}
\usepackage{caption}
\usepackage{bbding}
\usepackage{amsfonts,amsthm,array} 
\usepackage[ruled]{algorithm2e}
\usepackage{makecell}

\begin{document}
	
	
    \title{Trajectory and Power Design for Aerial CRNs with Colluding Eavesdroppers	\thanks{Manuscript received.}}

    \author{Hongjiang~Lei, 
    	Jiacheng~Jiang,
    	Haosi~Yang,
    	Ki-Hong~Park, \\
    	Imran~Shafique~Ansari, 
    	Gaofeng~Pan, 
    	and~Mohamed-Slim~Alouini 
	\thanks{This work was supported by the National Natural Science Foundation of China under Grant 61971080.}
	\thanks{H. Lei is with the School of Communications and Information Engineering, Chongqing University of Posts and Telecommunications, Chongqing 400065, China and also is Chongqing Key Lab of Mobile Communications Technology, Chongqing 400065, China (e-mail: leihj@cqupt.edu.cn).}
	\thanks{J.~Jiang, and H. Yang are with the School of Communications and Information Engineering, Chongqing University of Posts and Telecommunications, Chongqing 400065, China (e-mail: cquptjjc@163.com, cquptyhs@163.com).}
	\thanks{I. S. Ansari Bharath Institute of Science and Technology, Chennai 600073, India (e-mail: ansarimran@ieee.org).}
	\thanks{G. Pan is with the School of Cyberspace Science and Technology, Beijing Institute of Technology, Beijing 100081, China (e-mail: gaofeng.pan.cn@ieee.org).}
	\thanks{K.-H. Park and M.-S. Alouini are with CEMSE Division, King Abdullah University of Science and Technology (KAUST), Thuwal 23955-6900, Saudi Arabia (e-mail: kihong.park@kaust.edu.sa, slim.alouini@kaust.edu.sa).}
}

\maketitle
	
\begin{abstract}
Unmanned aerial vehicles (UAVs) can provide wireless access services to terrestrial users without geographical limitations and will become an essential part of the future communication system. However, the openness of wireless channels and the mobility of UAVs make the security of UAV-based communication systems particularly challenging. This work investigates the security of aerial cognitive radio networks (CRNs) with multiple uncertainties colluding eavesdroppers. A cognitive aerial base station transmits messages to cognitive terrestrial users using the spectrum resource of the primary users. All secondary terrestrial users and illegitimate receivers jointly decode the received message. The average secrecy rate of the aerial CRNs is maximized by jointly optimizing the UAV's trajectory and transmission power. An iterative algorithm based on block coordinate descent and successive convex approximation is proposed to solve the non-convex mixed-variable optimization problem. Numerical results verify the effectiveness of our proposed algorithm and show that our scheme improves the secrecy performance of airborne CRNs.
\end{abstract}
	

\section{Introduction}
\label{sec:introduction}

\subsection{Background and Related Works}
\label{sec:Background}

Due to their high mobility, low cost, and on-demand deployment flexibility, unmanned aerial vehicles (UAVs) are considered critical to the future of the Internet of Things (IoT), enabling the interconnection of all things \cite{ZengY2019Proc, LiB2019IoT}.
{More and more communication scenarios are now becoming flexible and diverse due to the inclusion of UAVs, such as UAV-assisted real-time surveillance, traffic control, and emergency disaster relief. In these application scenarios, UAV-assisted communication is considered to be an essential component of future mobile communication systems. }
By designing the position of UAVs or optimizing their flight trajectory, the air-to-ground (A2G) links between UAVs and terrestrial nodes are line-of-sight (LoS) with a high probability.
Therefore, the trajectory of UAVs has a decisive impact on the performance of the UAV-based communication system and becomes a fundamental problem to be solved in the design of airborne communication systems \cite{WuQ2019WC}.

{
	The concept of the ``Internet of Everything" has led to an increasingly prominent role for UAVs in wireless communication systems \cite{YanC2019Access}.
	UAV-assisted communication systems utilize drones to provide ubiquitous wireless coverage for a given area.
	In varying application scenarios, drones can be used in different roles in communication systems, such as base stations (BSs) \cite{WuQ2018TWC} - \cite{YangL2020IoT}, relays \cite{LiuT2021TGCN} - \cite{WangC2022TCOM}, and users \cite{ZhanC2018WCL} - \cite{LuoW2021IoT}.
	In \cite{WuQ2018TWC}, a communication system consisting of multiple aerial BS was studied.
	The UAV trajectory was optimized to maximize the minimum throughput under the consideration of user scheduling coefficient and UAV transmit power.
	An efficient iterative algorithm based on successive convex approximation (SCA) and block coordinate descent (BCD) was proposed for mixed integer non-convex optimization. Their results showed that the proposed algorithm could improve the system's utility.
	Considering the limited energy consumption of UAVs' flight and ensuring the system's service quality, the author in \cite{BejaouiA2020WCL} investigated the max-min fairness problem to maximize the throughput of the system by jointly optimizing the UAV transmit power and flight trajectory.
	In \cite{WangJ2019IoT}, a space-air-ground three-tier heterogeneous network consisted of a satellite, multiple UAVs, and multiple terrestrial BS to provide seamless coverage service for the ground users.	
	A two-stage joint optimization was proposed to deal with the intra- and inter-interference and cross-tier resource allocation among different networks. 
	A multi-UAVs-aided mobile edge computing system was designed to balance traffic and a new multi-UAV deployment mechanism was proposed in \cite{YangL2020IoT}. 
	A near-optimal solution algorithm was proposed to solve the load-balancing multi-UAV deployment problem.
When UAVs are used as relay nodes, long-distance wireless links between terrestrial BSs and terrestrial nodes (TNs) can be quickly established, which is helpful in disaster or other emergency scenarios \cite{ZengY2019Proc}.
In \cite{LiuT2021TGCN}, authors considered a UAV-assisted cooperative system in which multiple UAVs were utilized as decode-and-forward relays to forward signals to the TNs.
The minimum transmission rate of the system was maximized by jointly optimizing the flight trajectory of the relay UAVs, the transmit power on UAVs and terrestrial BS.
An iterative algorithm was proposed to solve the joint optimization problem based on the block coordinate ascent technique, introducing the slack variable and SCA techniques.
In \cite{LeeJH2021TCOM}, an aerial relay balanced the difference in transmission rates between free-space optical and radio frequency links.
The throughput for delay-limited and delay-tolerant scenarios was maximized by carefully designing the flight trajectory of the aerial relay, which in turn maximized the throughput.
Combining intelligent reflecting surface (IRS) and UAV can establish new channels or improve the A2G quality through full-angle reflection in the air.
A UAV-IRS-aided covert scheme was proposed in \cite{WangC2022TCOM}, where the UAV carried an IRS and was utilized as a relay. Considering Willie's imperfect position state information, the covert transmission rate was maximized by alternately optimizing Alice's transmit power, IRS phase shift, and the horizontal position of the UAV-IRS.
In UAV-aided wireless sensor networks and IoT communication systems, UAVs are utilized as aerial access points to disseminate/collect information to/from TNs \cite{ZengY2019Proc}.
A mixed-integer non-convex optimization problem was proposed to prolong the wireless sensor network lifetime and the UAV's trajectory was jointly optimized to minimize the maximum energy consumption of all sensor nodes in \cite{ZhanC2018WCL}.
Refs. \cite{LuW2021CIS, LuoW2021IoT} combined UAV trajectory design and wireless charging technology in the scenarios of the Internet of Things, considered the two-way link between information transmission and energy acquisition, and realized the problem of maximum and minimum data acquisition by optimizing the user scheduling coefficient and UAV flight trajectory.
}

The security of UAV communication applications is vulnerable to threats due to the broadcast nature and the prevailing LoS channel conditions \cite{WuQ2019WCPLS}, \cite{WangHM2019WC}, \cite{UllahZ2020TCCN}.
Since then, more and more researches have focused on the security of UAV-based communication systems.
The security of A2G and ground-to-air (G2A) links was investigated in \cite{ZhangG2019TWC}, and the average secrecy rate (ASR) were maximized by jointly optimizing the transmission power and the trajectory of the UAV.
By using successive convex approximation (SCA) and block coordinate descent (BCD), iterative algorithms were proposed to solve the non-convex problems.
In \cite{WangY2021TCCN}, the security of underlay cognitive radio networks (CRNs) was investigated.
The ASR was maximized by designing an antenna jammer's trajectory and transmit power.
However, the eavesdroppers were assumed to operate in half-duplex mode and their position was assumed to be perfectly known.

The full-duplex (FD) eavesdroppers pose more severe security threats to UAV-based communication systems because they can not only intercept confidential information from A2G links but also send jamming signals to degrade the reception quality of the legitimate link.
In \cite{DuoB2021ChinaCom}, authors analyzed the security risks caused by terrestrial FD eavesdroppers and propose an iterative algorithm to maximize the ASRs of the downlink and uplink transmissions by optimizing the UAV flight trajectory and transmission power.
The proposed non-convex problem was decomposed into sub-problems in BCD scheme with suboptimal solutions obtained by the SCA. 
The position of the eavesdroppers at the base stations was also assumed to be perfectly known.

{Estimating the exact location and channel state information in these scenarios is very challenging, especially for malicious eavesdroppers with some deception equipment.}
In these scenarios, the bounded eavesdropper's location error model was utilized and the worst ASR (WASR) was considered.
Ref. \cite{CuiM2018TVT} investigated the security of an A2G communication system with multiple eavesdroppers, whose locations are assumed to be imperfect. 
{Another UAV was utilized as a friendly jammer to enhance the security of the UAV-aided communication, Ref. \cite{ZhongC2019CL} designed both the aerial BS and the aerial jammer's transmit power and trajectories to maximize the WASR.}
The WASR and the secrecy energy efficiency (SEE) of the considered system were maximized by jointly optimizing the trajectory and transmit power of the aerial base station, respectively.
Considering the imperfect location of the eavesdroppers and the practical propulsion energy consumption, two cooperative dual UAV-assisted data collection schemes have been proposed in \cite{ZhangR2021TWC}.
The WASR and energy efficiency were maximized by optimizing the flight trajectory of the dual UAVs, the transmission power, and the scheduling coefficients.
Furthermore, in some practical scenarios, multiple eavesdroppers attempt to intercept confidential information.
In these scenarios, the base station needs to move away from all the eavesdroppers as far as possible and reach the legitimate targets as closely as possible to increase the secrecy rate.
Authors in \cite{WangW2021JSAC} studied the secrecy performance of the energy-constrained dual-UAV systems with multiple inaccurate location eavesdroppers.
The WASR was maximized by optimizing the 3D trajectories, the transmit power of the airborne transmitter and jammer, and the time slot allocation scheme of the airborne transmitter.
The security of an underlay CRN with multiple imprecise location eavesdroppers was investigated in \cite{ZhouYF2020TCOM}.
The ASR of the cognitive user was maximized by optimizing the UAV's trajectory and transmission power.
Authors in \cite{NguyenPX2021TVT} investigated the secure CRNs and a friendly jammer was utilized to transmit artificial noise.
Both scenarios were considered in which all the terrestrial nodes' locations were known and unknown.
{A dual collaborative underlay aerial IoT system with 
multiple uncertainty potential eavesdropper was considered and the WASR was maximized by jointly considering UAV trajectories, user scheduling, and transmit power in \cite{LeiH2023IoT}.}

To maximize the eavesdropped information in some scenarios, the eavesdroppers may engage in colluding behavior, i.e., gathering and sharing information, especially in those scenarios where the quality of the eavesdropping links is poor.
The most hazardous scenario is when all eavesdroppers are involved in malicious collusion \cite{LiuS2022TWC}.
The security of cellular UAV communication systems with multiple legitimate and illegitimate ground receivers was investigated in \cite{YaoJ2020TWC3D}.
The secrecy rate of the UAV communication systems was maximized for both non-colluding and colluding scenarios by jointly optimizing the UAVs' 3D location and transmit power. However, the location of all eavesdroppers was assumed to be perfectly known.
Authors in \cite{GaoY2021CL} studied the energy efficiency of a UAV-aided system with multiple colluding eavesdroppers with imperfect location information and formulated a mixed-integer non-convex optimization problem to minimize the energy consumption by jointly optimizing the UAV's trajectory and user scheduling.
Authors in \cite{FuH2022TVT} investigated the secrecy performance of a UAV-aided non-orthogonal multiple access (NOMA) systems with multiple colluding eavesdroppers with imperfect location information.
The minimum WASR was maximized by jointly optimizing the UAV's trajectory, the transmit power, and the power allocation coefficient.
{ Table \ref{table1} outlines recent works in literature related to trajectory design of UAV communication systems.}

\begin{table*}[tb]
	\centering
	{
	\caption{\textit{Related to Trajectory Design of UAV Communication Systems.}}
		\label{table1}
		\centering
		\begin{tabular}{|c|c|c|c|c|c|c|}
			\hline
			{Reference} &{Security}&{Role of UAVs} & {\makecell[c]{Imperfect Location \\ of Eavesdroppers}} &{\makecell[c]{Colluding \\Eavesdroppers}}& {FD Eavesdroppers} &{CRNs}\\
			\hline
			{ \cite{WuQ2018TWC},\cite{BejaouiA2020WCL}} &{}&{BS}&{}&{}&{}&{}\\
			\hline
			{\cite{LiuT2021TGCN}-\cite{	WangC2022TCOM}} &{}&{Relay}&{}&{}&{}&{}\\
			\hline
			{\cite{ZhanC2018WCL}-\cite{LuoW2021IoT}} &{}&{User}&{}&{}&{}&{}\\
			\hline
			{\cite{ZhangG2019TWC}} &\checkmark&{BS, User}&{}&{}&{}&\\
			\hline
			{\cite{WangY2021TCCN}} &\checkmark&{Jammer}&{}&{}&{}&\checkmark\\
			\hline
			{\cite{DuoB2021ChinaCom}}&\checkmark&{BS, User}&{}&{}&{\checkmark}&{}\\
			\hline
			{\cite{CuiM2018TVT}}&\checkmark&{BS} &{\checkmark}&{}&{}&{} \\
			\hline
			{\cite{ZhongC2019CL}}&{\checkmark}&{BS, Jammer}&{\checkmark}&{}&{}&{}\\			
			\hline
			{\cite{ZhangR2021TWC}}&\checkmark&{User, Jammer}&{\checkmark}&{}&{}&{}\\
			\hline
			{\cite{WangW2021JSAC}}&{\checkmark}&{BS, Jammer}&{\checkmark}&{}&{}&{}\\
			\hline
			{\cite{ZhouYF2020TCOM}}&{\checkmark}&{BS}&{\checkmark}&{}&{}&\checkmark\\
			\hline
			{\cite{NguyenPX2021TVT}}&{\checkmark}&{Jammer}&{\checkmark}&{}&{}&{\checkmark}\\
			\hline
			{\cite{LiuS2022TWC}}&{\checkmark}&{BS}&{}&{\checkmark}&{}&{}\\
			\hline
			{\cite{LeiH2023IoT}}&{\checkmark}&{BS, Jammer}&{\checkmark}&{}&{}&{\checkmark}\\
			\hline
			{\cite{YaoJ2020TWC3D}}&{\checkmark}&{BS}&{}&{\checkmark}&\checkmark&{}\\
			\hline
			{\cite{GaoY2021CL}}&{\checkmark}&{BS}&{\checkmark}&{\checkmark}&{}&{}\\
			\hline
			{\cite{FuH2022TVT}}&{\checkmark}&{BS}&{\checkmark}&{\checkmark}&{}&{}\\
			\hline
			{Our Work}&{\checkmark}&{BS}&{\checkmark}&{\checkmark}&{\checkmark}&{\checkmark}\\
			\hline
	\end{tabular}
}
\end{table*}

\subsection{Motivation and Contributions}
\label{sec:Motivation}

{
The previous discussed works demonstrated that the security of UAV-aided CRNs can be significantly improved by designing the UAV’s trajectory and other system parameters (such as transmit power).
However, these outstanding works did not answer the following security issue:
\textit{
	How to design the trajectory and transmit power of the aerial base station where multiple inaccurate location FD eavesdroppers intercept information in colluding mode?
	What is the optimal secrecy performance of the aerial CRNs with multiple colluded FD eavesdroppers with imperfect location?
	}
Hence, this work answers these questions by optimizing the WASR of aerial CRNs.
It is assumed that multiple location-uncertainty illegitimate receivers work in FD and colluding modes and the legitimate receivers jointly decode the received legitimate messages via maximal ratio combining (MRC) scheme.
}
The main contributions of this work are summarized as follows:

\begin{enumerate}
	\item We consider an underlay CRN consisting of an aerial base station and multiple primary and cognitive users.
	Multiple malicious FD eavesdroppers at uncertain locations wiretap confidential information and transmit interference to degrade the reception quality of the legitimate links. All terrestrial secondary users and illegitimate receivers are assumed to utilize the cooperative decoding of received message, respectively. 	
	The WASR is maximized by jointly optimizing the UAV's transmission power and trajectory while satisfying SEE limit.
	Due to its non-convexity, the optimization problem is split into two sub-problems and transformed into approximated convex forms via SCA.
	The BCD technique is utilized to solve these subproblems.
	
	\item The proposed algorithm's convergence and complexity are analyzed and the numerical results of the proposed scheme are compared with three benchmarks where the aerial base station works with the fixed trajectory and transmits power simultaneously or only optimizing the trajectory and transmit power separately. The efficiency and the convergence of the proposed algorithm are verified.
	
	\item Relative to \cite{WangY2021TCCN}, \cite{ZhouYF2020TCOM}, \cite{NguyenPX2021TVT,LeiH2023IoT} where the secure CRNs with a single eavesdropper was considered, this work investigates the security of aerial CRNs with multiple colluding eavesdroppers residing at uncertain location. Technically speaking, considering the collusion scenarios is much more challenging than the independent eavesdropping scenarios.
	
	\item Although the colluding scenarios were considered in \cite{LiuS2022TWC} and \cite{YaoJ2020TWC3D}, the results can not be applied to CRNs directly since the location of the eavesdroppers is assumed to be perfectly known and the underlay condition was not considered. This work assumes that both legitimate and illegitimate receivers are considered jointly to decode the received legitimate messages. Both the uncertainty of eavesdroppers' locations and the underlay condition are considered simultaneously.
	
\end{enumerate}

\subsection{Organization}
The rest of this paper is organized as follows.
The system model and problem formulation are provided in Section \ref{sec:SystemModel} and \ref{sec:Problem}, respectively.
A two-step alternating algorithm is proposed in Section  \ref{sec:ProposedAlgorithm1} to solve the problem.
Simulation results are demonstrated in Section \ref{sec:Simulation}.
Finally, Section \ref{sec:Conclusions} concludes this paper.
{ TABLE \ref{table2} lists the notations and symbols utilized in this work.}

\begin{table}[t]
	{
		\caption{\textit{List of Notations.}}
		\begin{center}
			\begin{tabular}{|c| c| }
				\hline
				\textbf{Notation}   	& \textbf{Description}								\\
				\hline
				${\mathbf{q}}_S^0$             & Initial location of $S$ 							\\
				\hline
				${\mathbf{q}}_S^F$               & Final location of $S$ 							\\
				\hline
				${{\mathbf{w}}_{{D_k}}}$         & Horizontal location of $D_k$ 					\\
				\hline
				${{\mathbf{w}}_{{E_m}}},{{\mathbf{\hat w}}_{{E_m}}}$     & Exact and estimated location of $E_m$  					\\
				\hline
				${{\mathbf{w}}_{{U_r}}}$			& Horizontal location of $U_r$ 					\\
				\hline
				${r_{{E_m}}}$                 & Maximized estimation errors of the distance 	\\
				\hline
				$H$  						  & Altitude of $S$ 								\\
				\hline
				${\Gamma _r}$               & Tolerance threshold of $U_r$  					\\
				\hline
				${\sigma ^2}$				& The noise power 							\\
				\hline
				$P_S^{\max }$				& The peak transmit power of $S$			\\
				\hline
				${P_{{E}}}$					& Transmit power of $E_m$				\\
				\hline
				$\alpha$					& Path loss exponent				\\
				\hline
				$V_S^{\max }$				& Maximum flight speed of the UAV		\\
				\hline
				${\beta _0}$				& Channel power gain at the reference distance		\\
				\hline
				${\delta _t}$				& Time slot length 								\\
				\hline
				$\varepsilon$				& Algorithm convergence precision					\\
				\hline
				$\Psi$				& SEE threshold of $S$			\\
				\hline	
			\end{tabular}
		\end{center}
		\label{table2}
	}
\end{table}

\section{System Model}
\label{sec:SystemModel}

\begin{figure}[t]
	\centering		
	\includegraphics[width = 2.5 in]{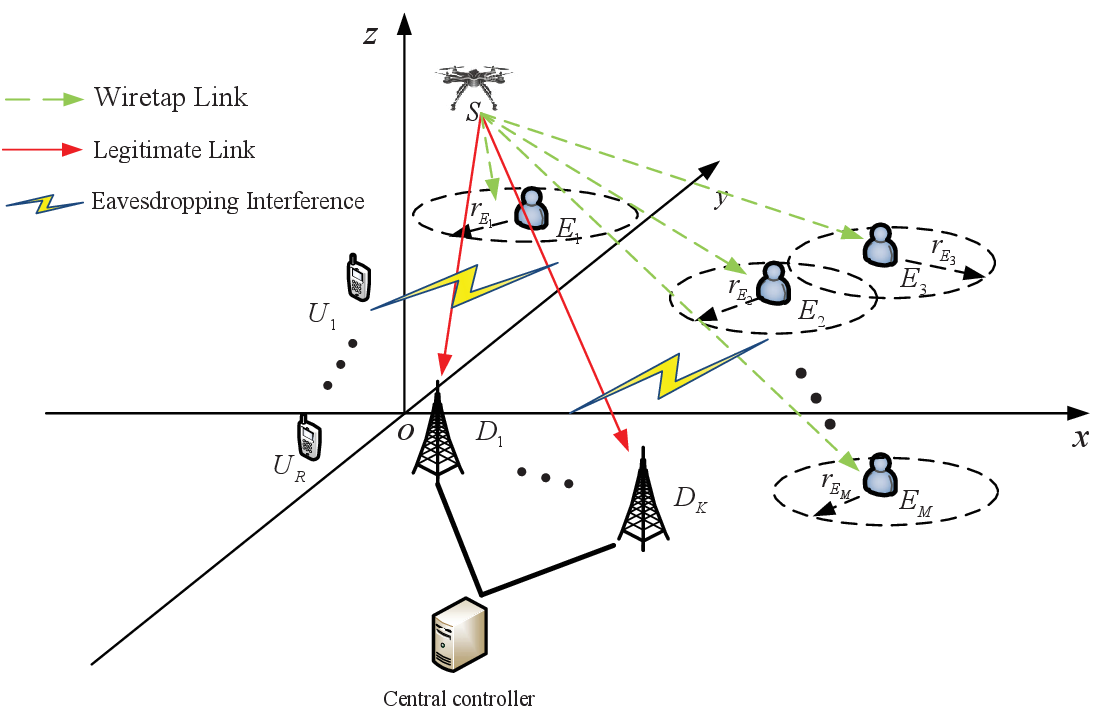}
    \caption{UAV-enabled CRN consists of an aerial base station (${S}$),  multiple primary users (${U_r}$), and multiple cognitive users (${D_k}$). Multiple FD eavesdroppers (${E_m}$) at uncertain locations not only try to wiretap the confidential information but also send interference to degenerate the reception quality of the legitimate link.}
    \label{fig_model}
\end{figure}

As shown in Fig. \ref{fig_model}, we consider an aerial underlay communication system consisting of a rotary-wing UAV that works as a base station (${S}$) and multiple terrestrial users (${D_k}, k = 1, \cdots, K$) in the presence of multiple primary users (${U_r}, r = 1, \cdots, R$) and multiple malicious FD eavesdroppers (${E_m}, m = 1, \cdots, M$).
All the nodes are equipped with a single antenna.

{ 
	To avoid collisions, $S$ flies at a constant altitude $H$ that is higher than the highest obstacle in the service area \cite{CuiM2018TVT}, \cite{ZhangR2021TWC}, \cite{ZhouYF2020TCOM}, \cite{LiuB2022TVT}.
	}
The flight period, $T$, is divided into ${N}$ time slots as ${\delta _t} = \frac{T}{N}$ \cite{WuQ2019WC}.
Without loss of generality, a 3D cartesian coordinate is used.
Then the coordinates of $S$, $D_k$, ${U_r}$, and $E_m$ are written as
${{\mathbf{q}}_{S}}\left( n \right) = {\left[ {{x_{S}}\left( n \right),{y_{S}}\left( n \right)} \right]^T}$,
${{\mathbf{w}}_{{D_k}}} = {\left[ {{x_{{D_k}}},{y_{{D_k}}}} \right]^T}$, 
${{\mathbf{w}}_{{U_r}}} = {\left[ {{x_{{U_r}}},{y_{{U_r}}}} \right]^T}$, and
${{\mathbf{w}}_{{E_m}}} {\left[ {{x_{{E_m}}},{y_{{E_m}}}} \right]^T}$, 
respectively, where
$n = 1, \cdots ,N$.

Similar to \cite{WangY2021TCCN} and \cite{ZhangR2021TWC},
the A2G links are assumed to be LoS links and the channel coefficient between $S$ and the terrestrial receiver $i$ is expressed as
\begin{equation}
	{h_{Si}}\left( n \right) = \frac{{{\beta _0}}}{{d_{Si}^2\left( n \right) + {H^2}}},
	\label{hSE}
\end{equation}
where
${i \in \left\{ {{D_k},{U_r}, {E_m}} \right\}}$,
${\beta _0}$ denotes the channel power gain at the reference distance and
${d_{Si}}\left( n \right) = \left\| {{{\mathbf{q}}_S}\left( n \right) - {{\mathbf{w}}_{i}}} \right\|$ signifies the Euclidean distance between $S$ and $i$, respectively.

Due to the uncertainty of the eavesdroppers' locations,
${{h_{SE}}}$ obtained by (\ref{hSE}) is imperfect.
We define ${{\mathbf{\hat w}}_{{E_m}}} \in {\mathbb{R}^{2 \times 1}}$ and ${r_{{E_m}}}$ to express the estimated location and { maximum error of the difference between the estimation and the practical distance, respectively.}
According to \cite{ZhangR2021TWC}, it is assumed that
$\left\| {{{\mathbf{q}}_S}\left( n \right) - {{{\mathbf{\hat w}}}_E}} \right\| \ge {r_E}$.
Following the triangle inequality, ${d_{S{E_m}}}$ is expressed as
\begin{equation}
	\begin{aligned}
		{d_{S{E_m}}}\left( n \right) &= \left\| {{{\mathbf{q}}_S}\left( n \right) - {{\mathbf{w}}_{{E_m}}}} \right\|\\
		&\ge \left| {\left\| {{{\mathbf{q}}_S}\left( n \right) - {{{\mathbf{\hat w}}}_{{E_m}}}} \right\| - \left\| {{{{\mathbf{\hat w}}}_{{E_m}}} - {{\mathbf{w}}_{{E_m}}}} \right\|} \right|\\
		&\ge \left| {\underbrace {\left\| {{{\mathbf{q}}_S}\left( n \right) - {{{\mathbf{\hat w}}}_{{E_m}}}} \right\|}_{ \buildrel \Delta \over = {{\hat d}_{S{E_m}}}\left( n \right)} - {r_{{E_m}}}} \right|,
		\label{dSE}	
	\end{aligned}	
\end{equation}
where ${{\hat d}_{S{E_m}}}\left( n \right)$ denotes the estimation of ${d_{S{E_m}}}\left( n \right)$.

The ground-to-ground (G2G) link between $D_k$ and $E_m$ is assumed to undergo quasi-static Rayleigh model, then channel coefficient is expressed as
\begin{equation}
	{h_{{D_k}{E_m}}} = \frac{{{{\beta _0}}}}{{{d_{{D_k}{E_m}}^\alpha }}}{\xi_{{D_k}{E_m}}},
	\label{hDE}	
\end{equation}
where ${d_{{D_k}{E_m}}} = \left\| {{{\mathbf{w}}_{{D_k}}} - {{\mathbf{w}}_{{E_m}}}} \right\|$ signifies the Euclidean distance between $D_k$ and $E_m$,
$\xi _{{D_k}{E_m}}$ follows an exponential distribution with unit mean,
and $\alpha  > 2$ denotes the path-loss exponent.
Similarly, the distance between $D_k$ and $E_m$ is expressed as
 \begin{equation}
 	\begin{aligned}
 		{d_{{D_k}{E_m}}} &= \left\| {{{\mathbf{w}}_{{D_k}}} - {{\mathbf{w}}_{{E_m}}}} \right\|\\
 		&\ge \left| {\left\| {{{\mathbf{w}}_{{D_k}}} - {{{\mathbf{\hat w}}}_{{E_m}}}} \right\| - \left\| {{{{\mathbf{\hat w}}}_{{E_m}}} - {{\mathbf{w}}_{{E_m}}}} \right\|} \right|\\
 		&\ge \left| {\underbrace {\left\| {{{\mathbf{w}}_{{D_k}}} - {{{\mathbf{\hat w}}}_{{E_m}}}} \right\|}_{ \buildrel \Delta \over = {{\hat d}_{{D_k}{E_m}}}} - {r_{{E_m}}}} \right|.
 		\label{dDE}	
 	\end{aligned}	
 \end{equation}
where $ {{\hat d}_{{D_k}{E_m}}} $ signifies the estimation of $	{d_{{D_k}{E_m}}}$.
Then ${{h_{SE_m}}}$ and ${h_{{D_k}{E_m}}}$ are approximated by
\begin{equation}
	{{\hat h}_{S{E_m}}}\left( n \right) = \frac{{{{\beta _0}}}}{{{{\left( {{{\hat d}_{S{E_m}}}\left( n \right) - {r_{{E_m}}}} \right)}^2} + {H^2}}},
	\label{hSE2}	
\end{equation}
and
\begin{equation}
	{{\hat h}_{{D_k}{E_m}}} = \frac{{{{\beta _0}}}}{{{{A_{k,m}}^\alpha }}}{\xi_{{D_k}{E_m}}}, 
	\label{hDE2}	
\end{equation}
respectively,
where
${A_{k,m}} = \left| {{{\hat d}_{{D_k}{E_m}}} - {r_{{E_m}}}} \right|$.

Due to the presence of FD eavesdropper nodes, the signal-to-interference-plus-noise ratio (SINR) at $D_k$ is expressed as
\begin{equation}
	{{\gamma _{{D_k}}}\left( n \right) = \frac{{{P_S}\left( n \right){h_{S{D_k}\left( n \right) }}}}{{\sum\limits_{i = 1}^M {{P_{{E_m}}}\left( n \right){\hat h_{{D_k}{E_m}} }}  + {\sigma ^2}}},}
	\label{gamma_D}
\end{equation}
where ${{P_S}\left( n \right)}$ and ${P_{{E_m}}}\left( n \right)$ denote the transmit power of the ${S}$ and $E_m$, respectively,
and ${{\sigma ^2}}$ signifies the variance of additive white Gaussian noise (AWGN).

With the coordinated multi-point (CoMP) reception, all the cognitive users jointly decode the received legitimate messages via maximal ratio combining (MRC) scheme \cite{YaoJ2020TWC3D}.
Then the achievable rate at the legitimate receiver is expressed as
{\begin{equation}
	\begin{aligned}
		{R_{S}}\left( n \right) & \approx  {{\mathbb{E}}  _\xi }\left[ {{{\log }_2}\left( {1 + \sum\limits_{k = 1}^K {{\gamma _{{D_k}}}} \left( n \right)} \right)} \right]\\
		&= {{\mathbb{E}}  _\xi }\left[ {{{\log }_2}\left( {1 + \sum\limits_{k = 1}^K {\frac{{{P_S}\left( n \right){h_{S{D_k}}}\left( n \right) }}{{\sum\limits_{m = 1}^M {{P_{{E_m}}}\left( n \right){\hat h_{{D_k}{E_m}}}}  + {\sigma ^2}}}} } \right)} \right],
		\label{rateS}
	\end{aligned}
\end{equation}
where 
${\mathbb{E}}  \left[  \cdot  \right]$ denotes the expectation.}

Since ${\hat h_{{D_k}{E_m}}}\left( n \right)$ in (\ref{hDE2}) is convex with respect to $\xi $,
${R_{S}}\left( n \right) $ in (\ref{rateS}) is convex with respect to  ${{\hat h}_{{D_k}{E_m}}}\left( n \right) $, based on Jensen's inequality,
we approximate the achievable at the legitimate receiver as \cite{HuaM2020TCOM}
\begin{equation}
		\begin{aligned}
			{R_{S}}\left( n \right) & \ge  {\log _2}\left( {1 + \sum\limits_{k = 1}^K {\frac{{{P_S}\left( n \right){h_{S{D_k}}\left( n \right) }}}{{\sum\limits_{m = 1}^M {{P_{{E_m}}}\left( n \right){{\mathbb{E}}  _\xi }\left[ {{\hat h_{{D_k}{E_m}}}} \right]}  + {\sigma ^2}}}} } \right)\\
			&= {\log _2}\left( {1 + \sum\limits_{k = 1}^K {\frac{{{P_S}\left( n \right){h_{S{D_k}}\left( n \right) }}}{{{\beta _0}}{\sum\limits_{m = 1}^M {{P_{{E_m}}}\left( n \right){{A_{k,m}^{ - \alpha }}}}  + {\sigma ^2}}}} } \right)\\
			& \buildrel \Delta \over = R_{S}^{\mathbf{L}}\left( n \right),
			\label{rateS2}
		\end{aligned}
\end{equation}
where the superscript `L' signifies the lower bound.

Similarly to (\ref{gamma_D}), the SINR at the $m$th $E$ can also be expressed as
\begin{equation}
	{{\gamma _{S{E_m}}}\left( n \right) = \frac{{{P_S}\left( n \right){{\hat h}_{S{E_m}}\left( n \right) }}}{{\sum\limits_{j \ne m} {{P_{{E_j}}}\left( n \right){h_{{E_j}{E_m}}}}  + {\sigma ^2}}},}
	\label{gamma_E}
\end{equation}
where ${{h_{{E_j}{E_m}}}}$ denotes the interference from other eavesdroppers.

It is assumed that all the eavesdroppers transmit signals with its maximum power over the whole period and cooperatively intercept/decode the confidential message from $S$.
Thus, we have ${P_{{E_m}}}\left( n \right) = {P_E}$.
Moreover, the worst-case scenario is considered, the interference from other eavesdroppers is assumed to
be perfectly cancelled \cite{DuoB2021ChinaCom}.
Therefore, we obtain the SINR at the illegitimate receiver as
{
	\begin{equation}
	\begin{aligned}
			{R_E}\left( n \right) &= {\log _2}\left( {1 + \sum\limits_{m = 1}^M {{\gamma _{{SE_m}}}} } \right)\\
			&={\log _2}\left( {1 + \sum\limits_{m = 1}^M {\frac{{{P_S}\left( n \right){{\hat h}_{S{E_m}}\left( n \right) }}}{{{P_E}\sum\limits_{j \ne m} {{h_{{E_j}{E_m}}}}  + {\sigma ^2}}}} } \right)\\
			&= {\log _2}\left( {1 + \sum\limits_{m = 1}^M {\frac{{{P_S}\left( n \right){{\hat h}_{S{E_m}}\left( n \right) }}}{{{\sigma ^2}}}} } \right).
	\end{aligned}
	\label{rateE}
\end{equation}
}
Then the achievable secrecy rate of the considered system is approximated as
\begin{equation}
	\begin{aligned}
		{R_{\sec }}\left( n \right) &={\left[ {{{\log }_2}\left( {R_S^{\mathbf{L}}\left( n \right) - R_E\left( n \right)} \right)} \right]^ + }\\
		&= {\left[ {{{\log }_2}\left( {\frac{{1 + \sum\limits_{k = 1}^K {\frac{{{P_S}\left( n \right){h_{S{D_k}}\left( n \right) }}}{{{{\beta _0}}{P_E}\sum\limits_{m = 1}^M {{{A_{k,m}^{ - \alpha }}}}  + {\sigma ^2}}}} }}{{1 + \sum\limits_{m = 1}^M {\frac{{{P_S}\left( n \right){{\hat h}_{S{E_m}}\left( n \right) }}}{{{\sigma ^2}}}} }}} \right)} \right]^ + },
		\label{RSEC1}
	\end{aligned}
\end{equation}
where ${\left[ x \right]^ + } = \max \left\{ {0,x} \right\}$. 
{
Then the SEE of the considered system is expressed as \cite{ZhangR2021TWC}
\begin{equation}
	\mu  = \frac{{\sum\limits_{n = 1}^N {{R_{\sec }}\left( n \right)} }}{{\sum\limits_{n = 1}^N {{P_S}\left( n \right)} }}.
	\label{SEE01}
\end{equation}
}

At the $n$ time slot, the interference at $U_r$ is expressed as
\begin{equation}
	\begin{aligned}
			{{\hat I}_r}\left( n \right) &= {P_S}\left( n \right){h_{S{U_r}}}\left( n \right) + {P_E}\sum\limits_{m = 1}^M {{\mathbb{E}_\xi }\left[ {{h_{{E_m}{U_r}}}} \right]}  \hfill \\
			&\mathop  \le \limits^{\left( a \right)} {P_S}\left( n \right){h_{S{U_r}}}\left( n \right) + {P_E}\sum\limits_{m = 1}^M {\frac{{{\beta _0}}}{{d_{{E_m}{U_r}}^\alpha }}}\\
			& \triangleq {I_r}\left( n \right) \hfill ,
			\label{IR1}
	\end{aligned}
\end{equation}
where ${h_{{E_m}{U_r}}} = \frac{{{\beta _0}}}{{d_{{E_m}{U_r}}^\alpha}}{\xi_{{E_m}{U_r}}}$ denotes the channel coefficient between ${E_m}$ and ${U_r}$, 
$\xi _{{E_m}{U_r}}$ follows an exponential distribution with unit mean,
{step ${\left( a \right)}$ is obtained based on Jensen's inequality \cite{BoydS2004Book}}
and ${d_{{E_m}{U_r}}} = \left| {\left\| {{{{\mathbf{w}}}_{{U_r}}}\left( n \right) - {{{\mathbf{\hat w}}}_{{E_m}}}} \right\| - {r_{{E_m}}}} \right|$.
To ensure the quality of service (QoS) of the primary users, the average interference caused by $S$ and $E_m$ must be limited to the interference temperature (IT) threshold, ${\Gamma _r}$,  \cite{WangY2021TCCN}, \cite{ZhouYF2020TCOM}, and \cite{NguyenPX2021TVT}.
Thus, we have
\begin{equation}
	\frac{1}{N}\sum\limits_{n = 1}^N {{I_r}\left( n \right)}  \le {\Gamma _r},\forall r.
	\label{IRconstraint}
\end{equation}

\section{Problem Formulation}
\label{sec:Problem}

This work optimizes the WASR of the aerial CRNs by designing $S$'s horizontal trajectory and transmission power, {while the SEE is guaranteed}.
Let ${\mathbf{Q}} = \left\{ {{{\mathbf{q}}_S}\left( n \right), \forall n} \right\}$, ${\mathbf{P}} = \left\{ {{P_{S}}\left( n \right),\forall n} \right\}$.
Thus, the optimization problem is formulated as
\begin{subequations}
	\begin{align}
			\mathcal{P}_{1} \,:\, &\mathop {\max }\limits_{{\mathbf{P}},{\mathbf{Q}}} \frac{1}{N}\sum\limits_{n = 1}^N {R_{\sec }^1\left( n \right)}  \label{P1a}\\
				{\mathrm{s.t.}}\; & {\mu  \ge \Psi,} \label{P1b}\\
				& 0 \le {P_S}\left( n \right) \le P_S^{\max },\forall n,\label{P1c}\\
				&\frac{1}{N}\sum\limits_{n = 1}^N {{I_r}\left( n \right)}  \le {\Gamma _r}, \forall r,   \label{P1d}\\
				&{{\mathbf{q}}_S}\left( 0 \right) = {{\mathbf{q}}_S^{\mathbf{I}}},{{\mathbf{q}}_S}\left( n \right) = {{\mathbf{q}}_S^{\mathbf{F}}},  \label{P1e}\\
				&\left\| {{{\mathbf{q}}_S}\left( n \right) - {{\mathbf{q}}_S}\left( {n - 1} \right)} \right\| \le V_S^{\max }{\delta _t}, \forall n,  \label{P1f}
	\end{align}
\end{subequations}
where 
{$\Psi$ is the SEE threshold,}
$P_S^{\max }$ is the peak transmit power of $S$,
${\mathbf{q}}_S^{\mathbf{I}}$ and ${\mathbf{q}}_S^{\mathbf{F}}$ signify the initial and final position of $S$, respectively,
$V_S^{\max }$ denotes the maximum speed  of $S$, 
(\ref{P1b}) is the SEE constraint,
(\ref{P1c}) is the power constraint,
(\ref{P1d}) is the constraint to ensure that the primary users work properly,
(\ref{P1e}) is the constraint for the initial and final position of UAVs,
and
(\ref{P1f})  is the constraint for the maximum flight path in adjacent periods of flight.

$\mathcal{P}_{1}$ is challengeable to solve since the operator ${\left[ x \right]^ + }$ makes the objective function non-smooth at zero value.
Moreover, $\mathcal{P}_{1}$ is a multivariate coupled optimization problem, which is non-convex concerning horizontal trajectory variables ${\mathbf{Q}} $ and the transmit power variables ${\mathbf{P}}$.

\section{Proposed Algorithm for Problem $\mathcal{P}_{1}$}
\label{sec:ProposedAlgorithm1}

{
To deal with the non-smoothness of ${R_{\sec }}\left( n \right)$, according to Lemma 1 in \cite{ZhangG2019TWC} and \cite{WangY2021TCCN}, removing the operator ${\left[ x \right]^ + }$ does not affect the optimal value of $\mathcal{P}_{1}$ because the secrecy rate can be obtained by setting the transmit power to zero when it is less than zero.
}
Simultaneously, based on the alternating optimization (AO) method,  ${{\mathbf{Q}}}$ and ${{\mathbf{P}}}$ are optimized in an alternating manner, taking into account the other given variables.

\subsection{Subproblem 1: Optimization Transmit Power of $S$ }
In this subsection, we optimize ${\mathbf{P}}$, the transmit power of $S$, with given ${\mathbf{Q}}$. The original problem $\mathcal{P}_{1}$ is rewritten as
\begin{subequations}
	\begin{align}
		\mathcal{P}_{1.1} \,:\, &\max \limits_{{{\mathbf{P}}}}\; \frac{1}{N}\sum\limits_{n = 1}^N {R_{\sec }^1\left( n \right)} \label{P11a}\\
					{\mathrm{s.t.}}\; &\;{\mu _1  \ge \Psi,} \label{P11b00}\\
					&\; (\textrm{\ref{P1c}}), (\textrm{\ref{P1d}}),
	\end{align}
\end{subequations}
where 
${\mu _1} = \frac{{\sum\limits_{n = 1}^N {R_{\sec }^1\left( n \right)} }}{{\sum\limits_{n = 1}^N {{P_S}\left( n \right)} }}$, 
$R_{\sec }^1\left( n \right) = {{\log }_2}\left( {1 + {A_n}{P_S}\left( n \right)} \right) - {{\log }_2}\left( {1 + {B_n}{P_S}\left( n \right)} \right)$,
${A_n} = \sum\limits_{k = 1}^K {\frac{{{h_{S{D_k}}\left( n \right) }}}{{{{\beta _0}}{P_E}\sum\limits_{m = 1}^M {{{A_{k,m}^{ - \alpha }}}}  + {\sigma ^2}}}} $,
and
${B_n} = \sum\limits_{i = 1}^M {\frac{{{\hat h_{S{E_m}}}\left( n \right) }}{{{\sigma ^2}}}} $. 

{
	To tackle the non-convex characterization in (\textrm{\ref{P11a}}) and (\textrm{\ref{P11b00}}), the first-order Taylor expansion is utilized to approximate  $R_{\sec }^1\left( n \right)$ as 
	\begin{equation}
		\begin{aligned}
			R_{\sec }^2\left( n \right) &= {{\log }_2}\left( {1 + {A_n}{P_S}\left( n \right)} \right) - {{\log }_2}\left( {1 + {B_n}P_S^{\left( l \right)}\left( n \right)} \right) \\
			&- \frac{{{B_n}}}{{\ln \left( 2 \right)\left( {1 + {B_n}P_S^{\left( l \right)}\left( n \right)} \right)}}\left( {{P_S}\left( n \right) - P_S^{\left( l \right)}\left( n \right)} \right),
			\label{RSEC2}
		\end{aligned}
	\end{equation}
	where 
	$P_S^{\left( l \right)}\left( n \right)$ is a given feasible point in the ${l}$th iteration.
	Then $\mathcal{P}_{1.1}$ is approximated as
	\begin{subequations}
		\begin{align}
			\mathcal{P}_{1.2} \,:\, &\max \limits_{{{\mathbf{P}}}}\; \frac{1}{{N}}\sum\limits_{n = 1}^N {R_{\sec }^2\left( n \right)} \label{019a}\\
			{\mathrm{s.t.}}\; &\; \sum\limits_{n = 1}^N {R_{\sec }^2\left( n \right)}  \ge \Psi \sum\limits_{n = 1}^N {{P_S}\left( n \right)}, \label{019b}\\
			&\; (\textrm{\ref{P1c}}), (\textrm{\ref{P1d}}). \label{019c}
		\end{align}
	\end{subequations}
	$\mathcal{P}_{1.2}$ is a convex problem and can be solved by existing optimization tools such as CVX.
}

\subsection{Subproblem 2: Optimizing Trajectory of $S$ }

Given ${\mathbf{P}}$, $\mathcal{P}_{1}$ is rewritten as
\begin{subequations}
	\begin{align}		
		\mathcal{P}_{2.1} \,:\, &\; \max\limits_{{{\mathbf{Q}}}}\; {\frac{1}{N}\sum\limits_{n = 1}^N {R_{\sec }^3\left( n \right)} } 	\label{P20a}\\
			{\mathrm{s.t.}}\; &\;{\mu _2  \ge \Psi,}  \label{P20b} \\
			&\; \frac{1}{N}\sum\limits_{n = 1}^N {I_r^{\mathbf{Q}}\left( n \right)}  \le {\Gamma _r}, \forall r, \label{P20c}\\
			&\; (\textrm{\ref{P1e}}), (\textrm{\ref{P1f}}),  \label{P20d}
	\end{align}
\end{subequations}
where 
{${\mu _2} = \frac{{\sum\limits_{n = 1}^N {R_{\sec }^3\left( n \right)} }}{{\sum\limits_{n = 1}^N {{P_S}\left( n \right)} }}$}, 
$R_{\sec }^3\left( n \right) = {\log _2}\left( {1 + \sum\limits_{k = 1}^K {\frac{{{f_{1,k}}}}{{d_{S{D_k}}^2 + {H^2}}}} } \right) $ $-  {\log _2}\left( {1 + \sum\limits_{m = 1}^M {\frac{{{f_2}}}{{{{\left( {{{\hat d}_{S{E_m}}} - {r_{{E_m}}}} \right)}^2} + {H^2}}}} } \right)$,
${f_{1,k}} = \frac{{{P_S}\left( n \right)}}{{{P_E}\sum\limits_{m = 1}^M {{A_{k,m}^{ - \alpha }}}  + \frac{1}{{{\rho _0}}}}}$,
${f_2} = {\rho _0}{P_S}\left( n \right)$,
$I_r^{\mathbf{Q}}\left( n \right) = \frac{{{P_S}\left( n \right){\rho _0}}}{{{d_{S{U_r}}^2} + {H^2}}} + {P_E}\sum\limits_{m = 1}^M {h_{{E_m}{U_r}}} $,
and
${\rho _0} = \frac{{{\beta _0}}}{{{\sigma ^2}}}$.

It must be noted that $\mathcal{P}_{2.1}$ is a non-convex problem because the objective function in (\ref{P20a}) is a non-concave function with respect to ${{{\mathbf{q}}_S}\left( n \right)}$ and (\ref{P20b}) and (\ref{P20c}) are non-convex constraints with respect to ${{{\mathbf{q}}_S}\left( n \right)}$.

By introducing new slack variables and using the successive convex approximation (SCA), we obtain
\begin{equation}
	\begin{aligned}
		{{R_{\sec }^3\left( n \right) }} &\ge {\log _2}\left( {1 + \sum\limits_{k = 1}^K {\frac{{{f_{1,k}}}}{{{\tilde \zeta _k}^{\left( l \right)}\left( n \right)}}} } \right) \\
		&- {\log _2}\left( {1 + \sum\limits_{m = 1}^M {\frac{{{f_2}}}{{{{\tilde \xi _m}}\left( n \right)}}} } \right)\\
		&- \frac{{\sum\limits_{k = 1}^K {\left( {\frac{{{f_{1,k}}}}{{{{\left( {{\tilde \zeta _k}^{\left( l \right)}\left( n \right)} \right)}^2}}}\left( {{{\tilde \zeta _k}}\left( n \right) - {\tilde \zeta _k}^{\left( l \right)}\left( n \right)} \right)} \right)} }}{{\ln \left( 2 \right)\left( {1 + \sum\limits_{k = 1}^K {\frac{{{f_{1,k}}}}{{{\tilde \zeta _k}^{\left( l \right)}\left( n \right)}}} } \right)}}\\
		&  \buildrel \Delta \over = R_{\sec }^{3,{\mathbf{L}}}\left( n \right),
	\end{aligned}
\end{equation}
where
the superscript `$l$' denotes the iteration index,
${{\tilde \zeta _k}}\left( n \right)$ and ${{\tilde \xi _m}}\left( n \right)$ are new slack variables, which must satisfy
\begin{equation}
	{{\tilde \zeta _k}}\left( n \right) \ge {d_{S{D_k}}^2} + {H^2}
	\label{20a}
\end{equation}
and
\begin{equation}
	{{\tilde \xi _m}}\left( n \right) \le {\left( {{{\hat d}_{S{E_m}}} - {r_{{E_m}}}} \right)^2} + {H^2},
	\label{20b}
\end{equation}
respectively.
Due to the right-hand-sight of (\ref{20b}) is convex which makes (\ref{20b}) become non-convex. According to SCA, (\ref{20b}) is rewritten as
\begin{equation}
	\begin{aligned}
			{{\tilde \xi _m}}\left( n \right) & \le {H^2} - 2{r_{{E_m}}}\left\| {{{\mathbf{q}}_S}\left( n \right) - {{\mathbf{w}}_{{E_m}}}} \right\| + r_{{E_m}}^2\\
			& + 2{\left( {{{\mathbf{q}}_S}\left( n \right) - {{\mathbf{w}}_{{E_m}}}} \right)^T}\left( {{{\mathbf{q}}_S}\left( n \right) - {\mathbf{q}}_S^{\left( l \right)}\left( n \right)} \right)\\
			&+ {\left\| {{\mathbf{q}}_S^{\left( l \right)}\left( n \right) - {{\mathbf{w}}_{{E_m}}}} \right\|^2}.
	\end{aligned}
\end{equation}

With the same method, (\ref{P20c}) is rewritten as
\begin{equation}
	\frac{1}{N}\sum\limits_{n = 1}^N {\left( {\left. {\frac{{{\rho _0}{P_S}\left( n \right)}}{{{{\tilde d}_{S{U_r}}} + {H^2}}} + {P_E}\sum\limits_{m = 1}^M {h_{{E_m}{U_r}}} } \right)} \right.}  \le {\Gamma _r},
\end{equation}
where ${{{\tilde d}_{S{U_r}}}}$ is a new slack variable which satisfies
\begin{equation}
	\begin{aligned}
		{{\tilde d}_{S{U_r}}} &\le {\left\| {{\mathbf{q}}_S^{\left( l \right)}\left( n \right) - {{\mathbf{w}}_r}} \right\|^2} \\
		&+ 2{\left( {{{\mathbf{q}}_S}\left( n \right) - {{\mathbf{w}}_r}} \right)^T}\left( {{{\mathbf{q}}_S}\left( n \right) - {\mathbf{q}}_S^{\left( l \right)}\left( n \right)} \right).
	\end{aligned}
\end{equation}

Based on the above conversion, $\mathcal{P}_{2.1}$ is approximated as
\begin{subequations}
	\begin{align}		
				{\mathcal{P}_{2.2}}&\mathop {\max }\limits_{{{\mathbf{q}}_S},{{\tilde \xi _m}}\left( n \right),{{\tilde \zeta _k}}\left( n \right),{{\tilde d}_{S{U_r}}}} {{\frac{1}{N}\sum\limits_{n = 1}^N {R_{\sec }^{3,{\mathbf{L}}}\left( n \right)} } } \\
				{\mathrm{s.t.}}\;  &\;{\mu _3  \ge \Psi,}  \\
				&\frac{1}{N}\sum\limits_{n = 1}^N {\frac{{{\beta _0}{P_S}\left( n \right)}}{{{{\tilde d}_{S{U_r}}} + {H^2}}}}  + {P_E}\sum\limits_{m = 1}^M {h_{{E_m}{U_r}}}  \le {\Gamma _r}, \forall r,\\
				& {{\tilde \zeta _k}}\left( n \right) \ge {d_{S{D_k}}^2} + {H^2}, \forall n, \forall k,\\
				&{{\tilde \xi _m}}\left( n \right) \le {H^2} - 2{r_{{E_m}}}\left\| {{{\mathbf{q}}_S}\left( n \right) - {{\mathbf{w}}_{{E_m}}}} \right\| + r_{{E_m}}^2	\nonumber	\\
				& \;\;\;\; \;\;\;\;  + 2{\left( {{{\mathbf{q}}_S}\left( n \right) - {{\mathbf{w}}_{{E_m}}}} \right)^T}\left( {{{\mathbf{q}}_S}\left( n \right) - {\mathbf{q}}_S^{\left( l \right)}\left( n \right)} \right)\\
				& \;\;\;\; \;\;\;\;  + {\left\| {{\mathbf{q}}_S^{\left( l \right)}\left( n \right) - {{\mathbf{w}}_{{E_m}}}} \right\|^2},\forall n,\forall m,		\nonumber 	\\
				&{{\tilde d}_{S{U_r}}} \le {\left\| {{\mathbf{q}}_S^{\left( l \right)}\left( n \right) - {{\mathbf{w}}_r}} \right\|^2}  \nonumber \\
				&\;\;  + 2{\left( {{{\mathbf{q}}_S}\left( n \right) - {{\mathbf{w}}_r}} \right)^T}\left( {{{\mathbf{q}}_S}\left( n \right) - {\mathbf{q}}_S^{\left( l \right)}\left( n \right)} \right), \forall n, \forall r, \\				
				& (\textrm{\ref{P1e}}),  (\textrm{\ref{P1f}}),
	\end{align}	
\end{subequations}
where 
${\mu _3} = \frac{{\sum\limits_{n = 1}^N {R_{\sec }^{3,{\mathbf{L}}}\left( n \right)} }}{{\sum\limits_{n = 1}^N {{P_S}\left( n \right)} }}$. 
Now $\mathcal{P}_{2.2} $ is convex and can be solved with existing optimisation tools.

\begin{table}[t]
	{
		\caption{\textit{List of Simulation Parameters.}}
		\begin{center}
			\begin{tabular}{|c |c| }
				\hline
				\textbf{Parameter}   				& \textbf{Value}\\
				\hline
				${\mathbf{q}}_S^0$              	& ${\left[ { -300,-200} \right]^T}$, ${\left[ { -250,800} \right]^T}$\\
				\hline
				${\mathbf{q}}_S^F$                	& ${\left[ { 100,-200} \right]^T}$, ${\left[ { 0,-200} \right]^T}$\\
				\hline
				${{\mathbf{w}}_{{D_k}}}$             &\makecell[c]{${\left[ { - 200,500; - 100,500;  0,500} \right]^T}$\\
																   ${\left[ { - 100,300; -  50,400;  0,300} \right]^T}$\\
																   ${\left[ { - 250,300; - 100,  0; 50,300} \right]^T}$}\\
				\hline
				${{\mathbf{\hat w}}_{{E_m}}}$        & \makecell[c]{${\left[ { -150,700; - 50,700} \right]^T}$\\
																	${\left[ { -125,100; -150,200} \right]^T}$\\
																	${\left[ { -200,-20;    0,-20} \right]^T}$}\\
				\hline
				${{\mathbf{w}}_{{U_r}}}$			& \makecell[c]{${\left[ { - 100,  0} \right]^T}$\\
																   ${\left[ { - 200,400} \right]^T}$\\
																   ${\left[ { - 100,400} \right]^T}$}\\
				\hline
				${r_{{E_m}}}$                 		&10 m, 30 m \\
				\hline
				$H$  								&100 m\\
				\hline
				${\Gamma _r}$               		& -110 dBm\\
				\hline
				${\sigma ^2}$						& -110 dBm\\
				\hline
				$P_S^{\max }$						&3 W\\
				\hline
				${P_{{E}}}$							&0.1 W\\
				\hline
				$\alpha$							&2.2\\
				\hline
				$V_S^{\max }$						& 60 m/s\\
				\hline
				${\beta _0}$						& - 60 dB\\
				\hline
				${\delta _t}$						&1 s\\
				\hline
				$\varepsilon$						&0.01\\
				\hline
				$\Psi$						&1\\
				\hline
			\end{tabular}
		\end{center}
		\label{table3}
	}
\end{table}

\subsection{Convergence Analysis of Algorithm 1}
Combining the above two subproblems and applying BCD, an iterative algorithm for solving $\mathcal{P}_{1}$ is proposed.
The suboptimal solution is obtained by selecting the initial feasible points and solving $\mathcal{P}_{1.2}$ and  $\mathcal{P}_{2.2}$ alternatively.
Each iteration's solutions serve as feasible input values at the next iteration.
The objective function of the original problem $\mathcal{P}_{1}$ at the $l$th iteration is defined as ${R_{\sec }}\left( {{{\mathbf{P}}^l},{\mathbf{Q}}^l} \right)$, meanwhile, the details of the entire iteration of $\mathcal{P}_{1}$ are summarized in Algorithm 1, where the tolerance of the convergence of the algorithm is denoted as $\varepsilon$.
The convergence of Algorithm 1 is proved as follows.
\begin{algorithm}[tb]
	\caption{Iterative Procedure of Problem $\mathcal{P}_{1}$}
	\KwIn{Initialization of feasible points.}
	\While
	{${R_{\sec }}\left( {{{\mathbf{P}}^{\left( l +1\right)}},{\mathbf{Q}}^{\left( l +1\right)}} \right) - {R_{\sec }}\left( {{{\mathbf{P}}^{\left( l \right)}},{\mathbf{Q}}^{\left( l \right)}} \right)  \ge  {\varepsilon }$}
	{1. Solve ($\mathcal{P}_{1.2}$) for given ${{\mathbf{Q}}^{\left( l \right)}}$ and obtain the solution ${{\mathbf{P}}^{{\left( l + 1 \right)}}}$;\\
		2. Solve ($\mathcal{P}_{2.2}$) for given $ {{{\mathbf{P}}^{{\left( l +1\right)}}}} $ and obtain the solution ${{\mathbf{Q}}^{{\left( l + 1 \right)}}}$;\\
		3. $l = l + 1$;\\
		4. Calculate ${R_{\sec }}\left( {{{\mathbf{P}}^{\left( l + 1 \right)}},{{\mathbf{Q}}^{\left( l +1\right)}}} \right)$.
	}
	\KwOut{${R_{\sec }}\left( {{{\mathbf{P}}^{\left( l +1 \right)}},{{\mathbf{Q}}^{\left( l + 1\right)}}} \right)$ with ${{\mathbf{P}}^*} = {{\mathbf{P}}^{\left( l +1\right)}},\;{{\mathbf{Q}}^*} = {{\mathbf{Q}}^{\left( l +1\right)}}$. }
\end{algorithm}

\begin{proof}
	
In Step 1 of \textbf{Algorithm 1}, the power allocation of UAV solved by applying interior point method. Thus, we have
\begin{equation}
	\begin{aligned}
		{R_{\sec }}\left( {{{\mathbf{P}}^{\left( l \right)}},{{\mathbf{Q}}^{\left( l \right)}}} \right) & \le {{R_{\sec }^{\mathbf{L},1}}}\left( {{{\mathbf{P}}^{\left( l + 1 \right)}},{{\mathbf{Q}}^{\left( l \right)}}} \right)\\
		&\le {R_{\sec }}\left( {{{\mathbf{P}}^{\left( l + 1 \right)}},{{\mathbf{Q}}^{\left( l \right)}}} \right),
	\end{aligned}
	\label{63}
\end{equation}
where the objective function of problem $\mathcal{P}_{1.2}$ is lower-bounded by ${{R_{\sec }^{\mathbf{L},1}}}\left( {{{\mathbf{P}}^{\left( l + 1 \right)}},{{\mathbf{Q}}^{\left( l \right)}}} \right)$.
Then, by solving $\mathcal{P}_{2.2}$ the suboptimal solution ${{{\mathbf{Q}}^{\left( l + 1 \right)}}}$ is obtained, and have
\begin{equation}
	\begin{aligned}
			{R_{\sec }}\left( {{{\mathbf{P}}^{\left( l + 1 \right)}},{{\mathbf{Q}}^l}} \right) & \le {{R_{\sec }^{\mathbf{L},2}}}\left( {{{\mathbf{P}}^{\left( l + 1 \right)}},{{\mathbf{Q}}^{\left( l + 1 \right)}}} \right)\\
			&\le {R_{\sec }}\left( {{{\mathbf{P}}^{\left( l + 1 \right)}},{{\mathbf{Q}}^{\left( l + 1 \right)}}} \right),
	\end{aligned}
	\label{63}
\end{equation}
where the objective function of problem $\mathcal{P}_{2.2}$ is lower-bounded by ${{R_{\sec }^{\mathbf{L},2}}}\left( {{{\mathbf{P}}^{\left( l + 1 \right)}},{{\mathbf{Q}}^{\left( l + 1 \right)}}} \right)$. 

The objective function is always non-decreasing after each iteration, as shown in (\ref{63}). Meanwhile, we additionally notice that due to the feasible set under the constraints, the objective function in $\mathcal{P}_{1}$ is upper bounded by a finite value. Therefore, Algorithm 1 is convergent.
\end{proof}

{
	The complexity of \textbf{Algorithm 1} comes from two aspects. 
	One is optimizing transmit power of $S$, the other is solving $S$'s trajectory. 
	They are solved by the interior point method with complexity ${\cal O}\left( {{\phi ^{3.5}}\ln \left( {\frac{1}{\varsigma  }} \right)} \right)$ and 
	${\cal O}\left( {{\varphi ^{3.5}}\ln \left( {\frac{1}{\varsigma  }} \right)} \right)$, respectively, 
	where $\phi  = 2N + R + 1$, ${\varphi  = NK + NM + NR + N + R + 5}$, and $\varsigma $ denotes convergence precision.
	It is assumed that the number of iterations is $I$. 
	Thus, the total complexity of \textbf{Algorithm 1} is ${\cal O}\left( {I\ln \left( {\frac{1}{\varsigma  }} \right)\left( {{\phi ^{3.5}} + {\varphi ^{3.5}}} \right)} \right)$.
}

\section{Numerical Results and Analysis}
\label{sec:Simulation}

\begin{figure*}[t]
	\centering
	\subfigure[The optimal trajectory of $S$.]{
		\label{fig2a}
		\includegraphics[width = 0.3 \textwidth]{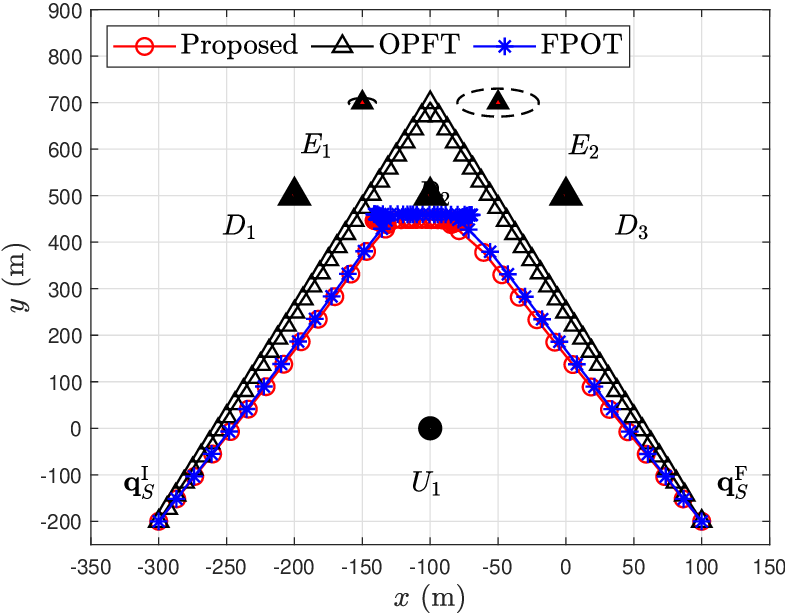}}
	\subfigure[The transmit power of $S$.]{
		\label{fig2b}
		\includegraphics[width = 0.3 \textwidth]{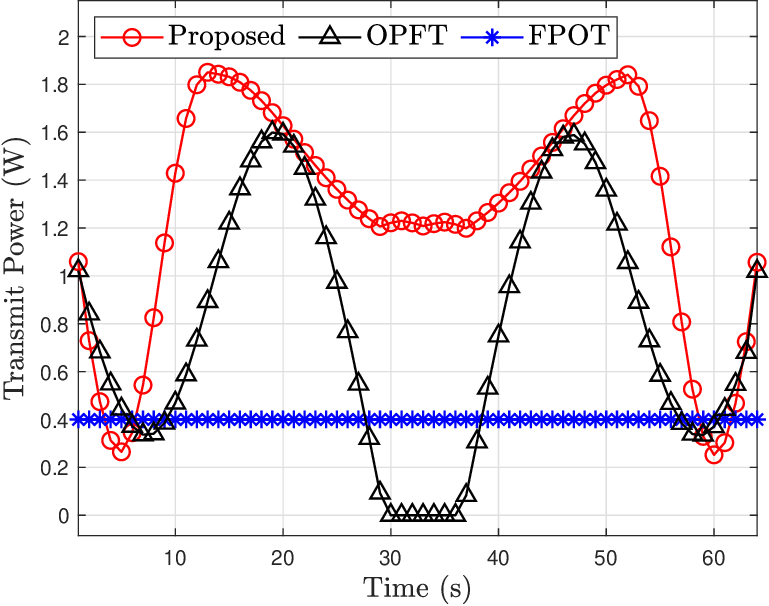}}
	\subfigure[The achievable secrecy rate.]{
		\label{fig2c}
		\includegraphics[width = 0.3 \textwidth]{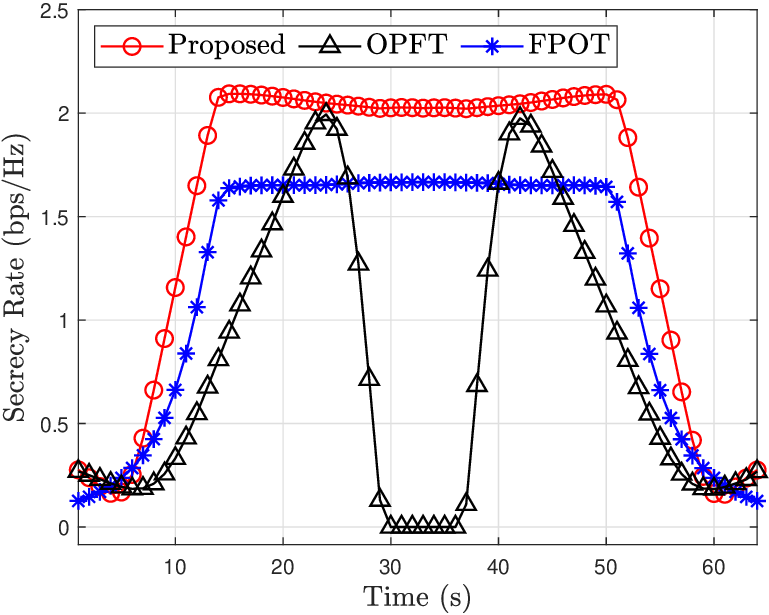}}
	\caption{Scenario 1 wherein the  FD-eavesdroppers are uniformly distributed with $K = 3$ terrestrial users.}
	\label{fig2}
\end{figure*}
\begin{figure*}[t]
	\centering
	\subfigure[The optimal trajectory of $S$.]{
		\label{fig03a}
		\includegraphics[width = 0.3 \textwidth]{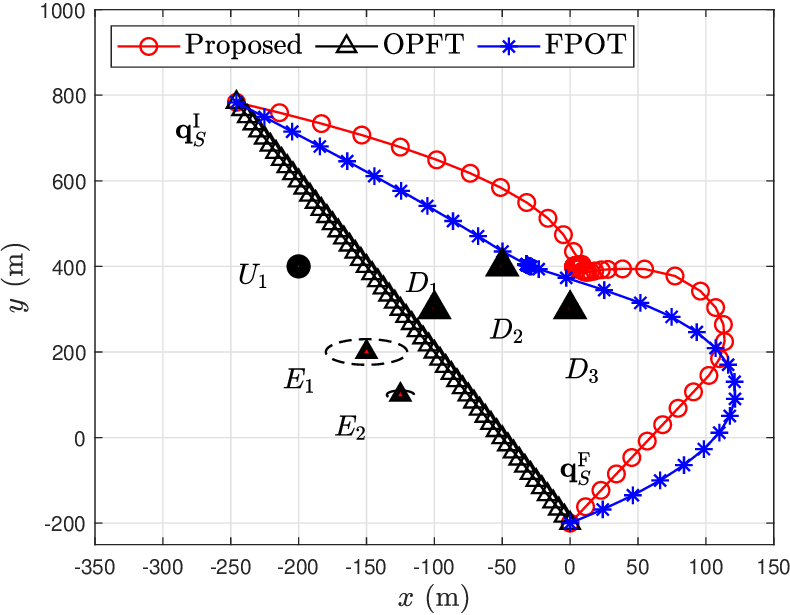}}
	\subfigure[The transmit power of $S$.]{
		\label{fig03b}
		\includegraphics[width = 0.3 \textwidth]{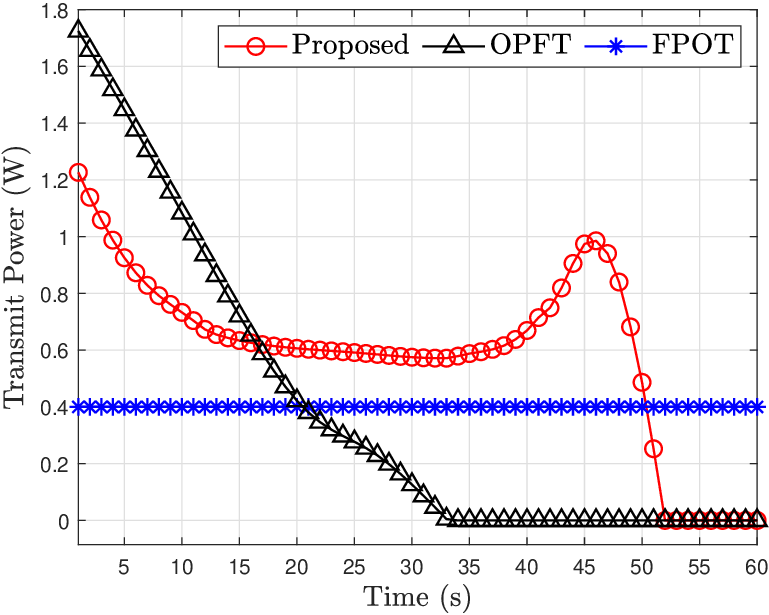}}
	\subfigure[The achievable secrecy rate.]{
		\label{fig03c}
		\includegraphics[width = 0.3 \textwidth]{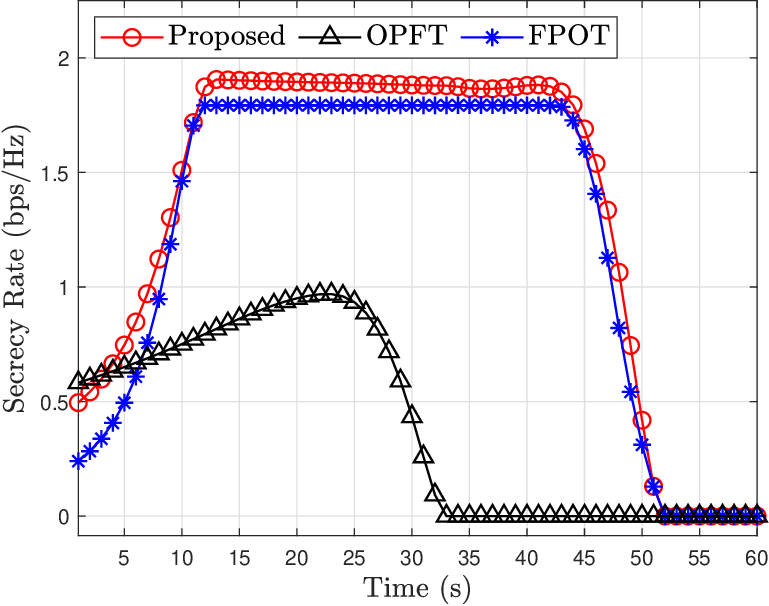}}
	\caption{Scenario 2 wherein the  FD-eavesdroppers are cluster distributed with $K = 3$ terrestrial users.}
	\label{fig03}
\end{figure*}
\begin{figure*}[t]
	\centering
	\subfigure[The optimal trajectory of $S$.]{
		\label{fig04a}
		\includegraphics[width = 0.3 \textwidth]{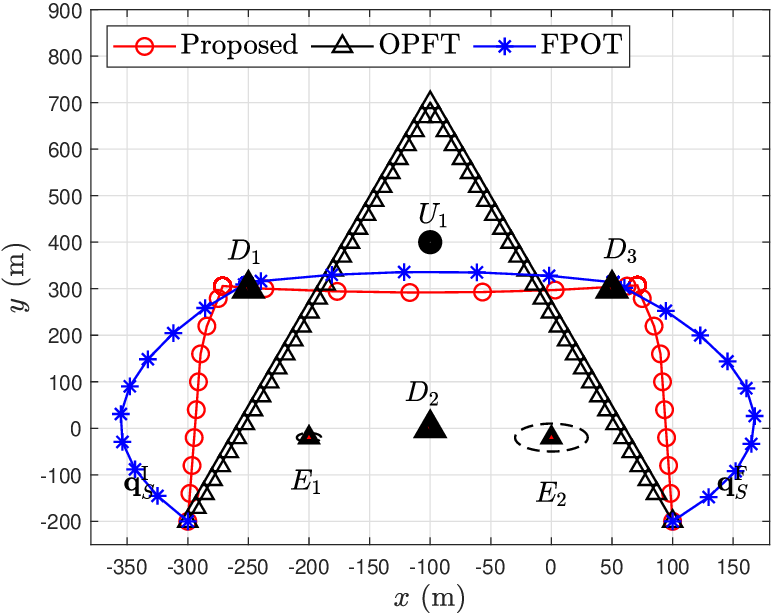}}
	\subfigure[The transmit power of $S$.]{
		\label{fig04b}
		\includegraphics[width = 0.3 \textwidth]{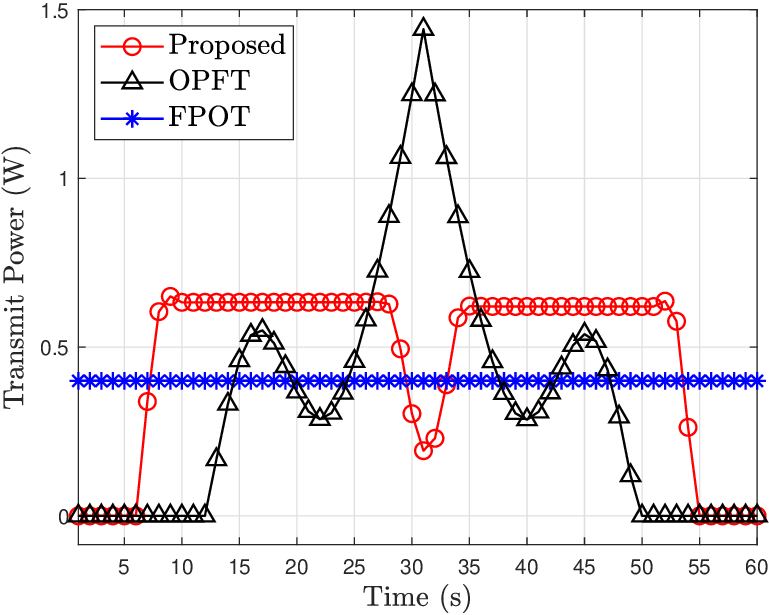}}
	\subfigure[The achievable secrecy rate.]{
		\label{fig04c}
		\includegraphics[width = 0.3 \textwidth]{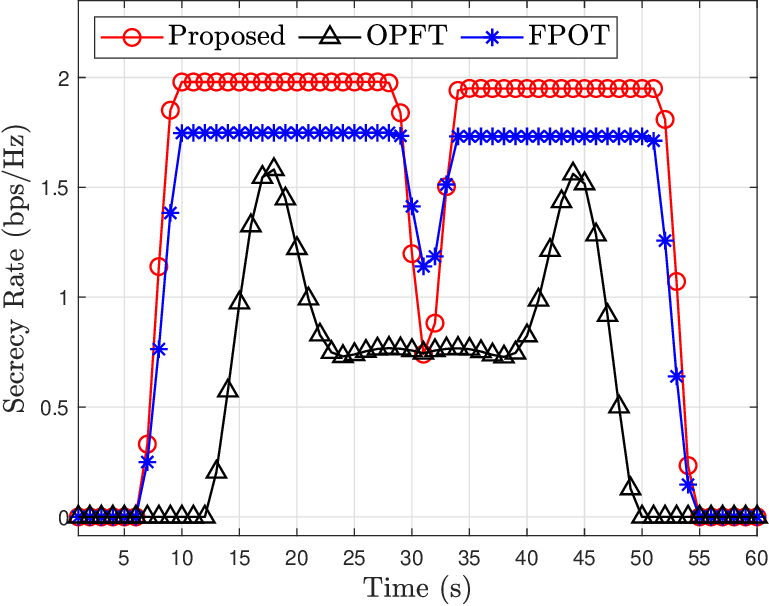}}
	\caption{Scenario 3 wherein the  FD-eavesdroppers are { interweaved in space} with $K = 3$ terrestrial users.}
	\label{fig04}
\end{figure*}

In this section, simulation results are presented to verify the performance of the joint transmit power control and UAV trajectory optimization algorithm. 
{To verify the effectiveness of our proposed algorithm, three scenarios are considered, in which the two FD-eavesdroppers are uniformly distributed, cluster distributed, and interweaved distributed with three terrestrial users, respectively.}
The details of parameter setup are listed in Table \ref{table3}.
Similar to \cite{ZhangG2019TWC}, \cite{WangY2021TCCN}, \cite{DuoB2021ChinaCom}, the following benchmark schemes are also considered :
1) Benchmark 1: $S$ operates with a fixed transmission power and a fixed trajectory, which is the initial value for the optimized transmit power and trajectory {(denoted as FPFT)}.
2) Benchmark 2: $S$ works with fixed transmit power and the trajectory is optimized {(denoted as FPOT)}.
3) Benchmark 3: $S$ works with fixed trajectory and its transmission power is optimized {(denoted as OPFT)}.
	
To illustrate the effect of colluding FD eavesdroppers, the trajectory obtained with different schemes for the different scenarios is shown in Figs. \ref{fig2} - \ref{fig04}.
For the { FPOT scheme} where $S$ works with fixed power, it should try to stay away from the area where the eavesdroppers are located while getting close to the legitimate users.
For the { OPFT scheme} where $S$ flies with a given trajectory, the transmit power should be reduced (to zero) as it approaches the area where the eavesdroppers are located to avoid the information being eavesdropped.
In the proposed scheme, the power is gradually increased/decreased to maximize the secrecy rate as it approaches the legitimate users (such as $D_1$ in Figs. \ref{fig2a} and \ref{fig04a}) and {ensure that the interference to the primary user does not exceed the IT threshold.}
As $S$ flies, it gradually moves away from the legitimate users and approaches the area where the eavesdroppers are located, and the transmission power is reduced to avoid the information being wiretapped. The characteristics of FPOT and OPFT are considered simultaneously to obtain the optimal secrecy performance. 
{It can be observed from Fig. \ref{fig04c} that the achievable secrecy rate of FPOT is higher than that of the proposed at 31s. 
	This is because the optimized transmit power of the proposed is lower to decrease the interference to the primary user.	
	We note that the optimality of the proposed scheme does not mean the proposed scheme is time-piecewise optimal but time-averaged optimal.  }

\begin{figure}[t]
	\centering		
	\includegraphics[width = 3in]{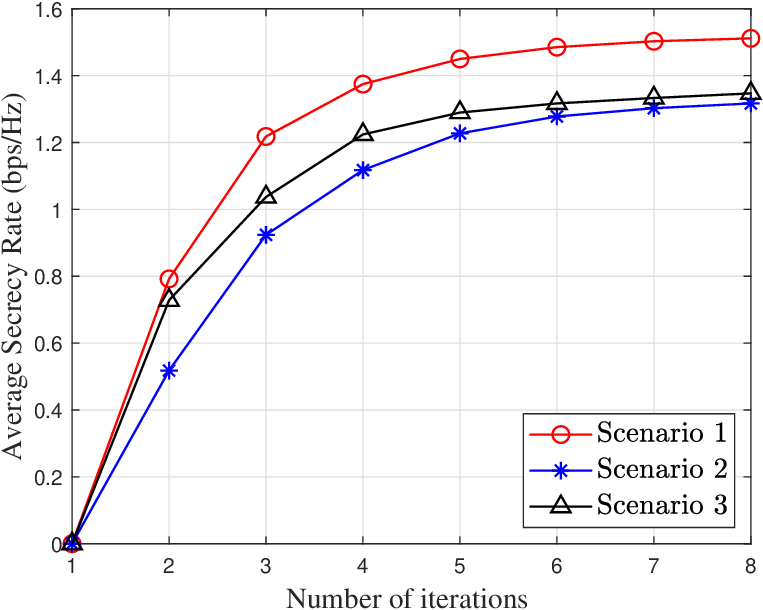}
	\caption{The WASR versus the number of iterations for different scenarios.}
	\label{Fig05iterations}
\end{figure}
Fig. \ref{Fig05iterations} shows the convergence of the proposed scheme in terms of the WASR versus the number of iterations in different scenarios.
The results illustrate the convergence of the proposed scheme in terms of jointly optimized transmit power and trajectory for the aerial base station.
One can observe that the WASR increases quickly within a small number of iterations and converges within around eight iterations in the listed three scenarios.	

\begin{figure}[t]
	\centering		
	\includegraphics[width = 3in]{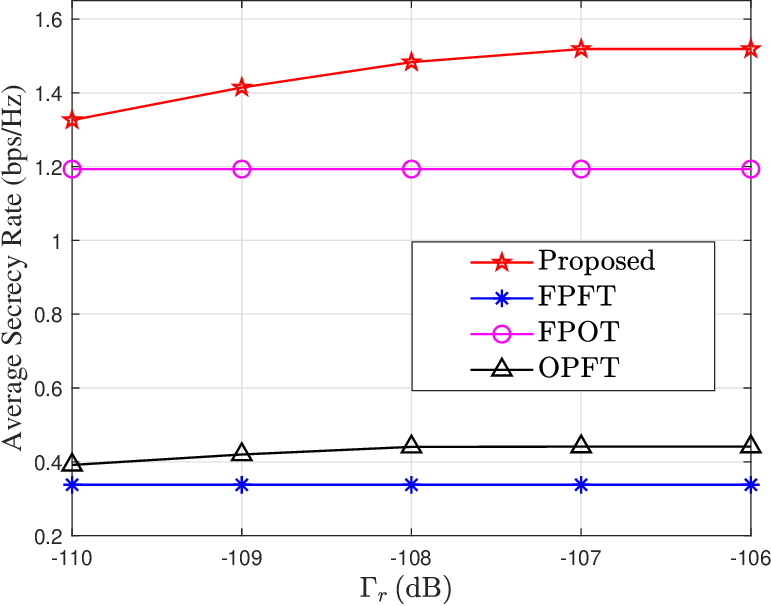}
	\caption{The WASR versus ${\Gamma _r}$ with different schemes.}
	\label{Fig06underlay}
\end{figure}
Fig. \ref{Fig06underlay} shows the effect of the IT threshold on the WASR of different schemes.
It can be observed that, as the threshold increases, the curves corresponding to the proposed and {OPFT} schemes show an increasing trend while those to  {FPFT and FPOT schemes} remain constant.
The reason is as ${\Gamma _r}$ increases, the optimized transmit power increases, allowing the secrecy rate of the system to be improved.
In the lower-${\Gamma _r}$ region, the transmit power of $S$ is lower since the average interference power caused by $S$ and eavesdroppers must be limited to the IT threshold to ensure that the QoS of the primary users is not affected. 
Thus, the secrecy rates with the proposed scheme and {FPOT scheme} outperform those with {FPFT and OPFT schemes}.
Moreover, the curves for the proposed and { OPFT schemes} have the apparent increasing trend because ${\Gamma _r}$ directly affects the transmit power of $S$ since the transmit powers in the proposed and {OPFT schemes} are adaptively changed while the transmit power in {FPFT and FPOT schemes} are fixed.
When the IT threshold reaches a certain value, the system becomes a non-cognitive mode where the transmit power is limited by the maximum transmit power of $S$.
Then the WASR is independent of the IT threshold in the larger-${\Gamma _r}$ region.

\begin{figure}[t]
	\centering		
	\includegraphics[width = 3in]{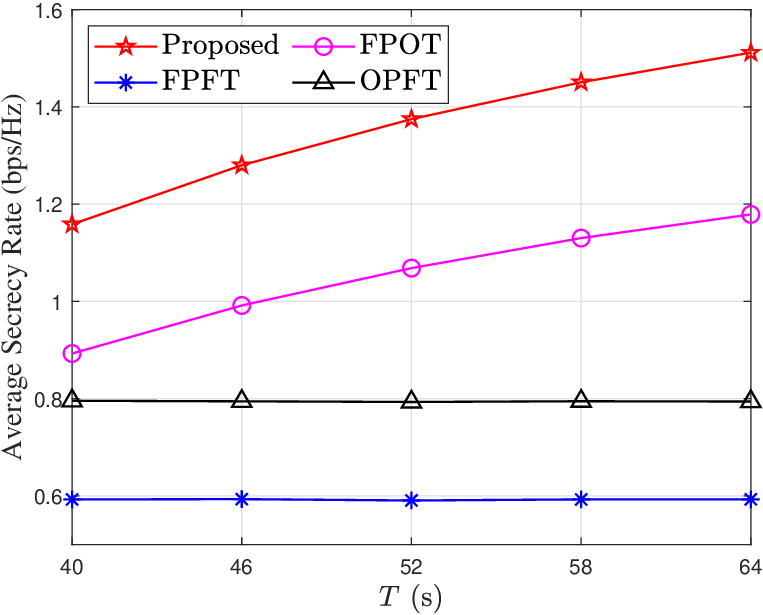}
	\caption{The WASR of scenario 1 under different schemes versus $T$.}
	\label{fig07_benchmark}
\end{figure}
Fig. \ref{fig07_benchmark} indicates the WASR versus the flight period $T$ for the considered system under different schemes.
It can be observed that the WASR of the considered system obtained by the proposed scheme outperforms those of others schemes.
This is because both the UAV's trajectory and transmission power are jointly optimized in the proposed scheme simultaneously.

\section{Conclusion}
\label{sec:Conclusions}

This work investigated the security of aerial CRNs with multiple uncertainty FD eavesdroppers.
By jointly designing the aerial base station's trajectory and transmission power, the WASR is formulated as a non-convex maximization problem.
Based on BCD and SCA technologies, a new efficient iterative algorithm is proposed to solve the non-convex problem, and suboptimal solutions are obtained.
Numerical results verified the convergence and effectiveness of our proposed algorithm. 
{ The A2G links in this work are assumed to be LoS links. 
In some scenarios, the A2G links are LoS links with elevation angle-dependent and environment-dependent probability.
Thus, designing the 3D trajectory will be an interesting problem in the future work.
Considering the distribution of the eavesdroppers' location is also an interesting work and will be part of the future work. 
}

\end{document}